\documentclass[11pt,reqno,makeidx]{amsart}

\usepackage[active]{srcltx}
\usepackage{amssymb,color}
\usepackage{float}
\restylefloat{table}

\usepackage[utf8]{inputenc}
\usepackage[T1]{fontenc}   
\usepackage[english]{babel} 
\usepackage{fullpage}
\usepackage{subfig}

\usepackage{amssymb,libertine}
  
\usepackage{array}
\usepackage{bbm}

\usepackage{mathtools}
\usepackage[svgnames]{xcolor}
\usepackage{hyperref}

\usepackage{wrapfig}
\usepackage{tikz}
\usepackage{color}
\usepackage{graphicx}
\usepackage{amssymb}
\usepackage{epstopdf}
 \usepackage{latexsym}
\usepackage{amsmath}
\usepackage{amsfonts} 
\usepackage{amsthm}
\usepackage[export]{adjustbox}

\usepackage{amssymb,libertine}
  
\usepackage{array}
\usepackage{bbm}

\usepackage{mathtools}
\usepackage[svgnames]{xcolor}
\usepackage{hyperref}

\usepackage{wrapfig}
\usepackage{tikz}
\usepackage{color}
\usepackage{graphicx}
\usepackage{amssymb}
\usepackage{epstopdf}
 \usepackage{latexsym}
\usepackage{amsmath}
\usepackage{amsfonts} 
\usepackage{amsthm}
\usepackage[export]{adjustbox}

\usepackage{ragged2e}
\usepackage{array}
\usepackage[active]{srcltx}
\usepackage{epsfig}
\usepackage{hyperref}
\usepackage{amsmath}
\usepackage{amssymb}
\usepackage{hyperref}
\usepackage{enumerate}

\newtheorem{Proposition}{Proposition}

  \newtheorem{Lemma}[Proposition]{Lemma}
  
  \newtheorem{Theorem}[Proposition]{Theorem}
 
 \newtheorem{Definition}[Proposition]{Definition}

 \def\RR{\mathbb{R}}

\def\Re{\mathrm{Re\,}}

\def\be{\begin{equation}}
\def\bel{\begin{equation}\label}
\def\ee{\end{equation}}

\def\bd{\begin{Definition}}
\def\ed{\end{Definition}}

\def\bp{\begin{Proposition}}
\def\bpl{\begin{Proposition}\label}
\def\ep{\end{Proposition}}
\def\bl{\begin{Lemma}}
\def\el{\end{Lemma}}
\def\bt{\begin{Theorem}}
\def\et{\end{Theorem}}

\def\om{\omega}

\def\ep{\epsilon}

\def\Ge{\geqslant}

\title{Ionization by an Oscillating Field: Resonances and Photons}
\author{Ovidiu Costin$^1$, Rodica D. Costin$^1$ \and Joel L. Lebowitz$^2$}

\date{%
$^1$The Ohio State University, 231 W 18th Ave, Columbus, OH 43210, email: costin.9@osu.edu\\%
$^2$Departments of Mathematics and Physics, Rutgers University, Hill Center - Busch Campus, 
110 Frelinghuysen Road Piscataway, NJ 08854, email:  lebowitz@math.rutgers.edu\\%
\today
}

\begin{document}
\maketitle


{\em Keywords:} ionization, oscillatory field, resonances, photons, model system, exact results

\begin{abstract}
We describe new exact results for a model of ionization of a bound state, induced by an oscillating potential. In particular we have obtained exact expressions, in the form of readily computable rapidly convergent sums, for 
the energy distribution of the emitted particles as a function of time, frequency and strength of the oscillating potential. Going beyond perturbation theory, these show resonances in the energy distribution which look like single or multi-photon absorption, similar to those observed in laser induced electron emission from solids or atoms. This is particularly so when the strength of the oscillating potential is small compared to the binding energy but is still visible for large fields, and even for time-periods of a few oscillations. We have also obtained the space-time structure of the wave function. Our model exhibits a form of stabilization; the ionization probability is not monotone in the strength of the oscillating potential.
\end{abstract}

{\bf Introduction.}
When (laser) light of frequency $\omega$ shines on a metallic surface or on a gas of atoms one observes the emission of electrons. 
This photo-electric effect is generally described (to leading order) along the lines in which Einstein first explained the phenomena in 1905 \cite{a}, \,\cite{b1}: an electron absorbs $n$ photons, thought of as "localized light particles", acquiring a kinetic energy $K=n\hbar\om-E_{b}$, where  $E_{b}$ is the minimum energy necessary to eject the bound electron from the metal or atom.

There is generally no specification of how the localized photons in a laser beam interact with and get absorbed by electrons. For this, one has presumably to go to a strongly coupled relativistic quantum field theory where the electro-magnetic field and its interaction with bound and free charges is properly described; see however \cite{c},\,\cite{j},\,\cite{d}. In practice one does computations "semi-classically" \cite{e},\,\cite{o1}. That is, one considers the electromagnetic field produced by the laser as a continuum non-quantized field. This simplification is considered plausible due to the large number of photons in a macroscopic field. 

The semi-classical description is given by 
  a non-relativistic Schr\"odinger equation $i\psi_t=(H_0+V(x,t))\psi$. Here $H_0$ describes the Hamiltonian of the reference system, e.g. a hydrogen atom, with both bound and free states
without the laser field and $V(x,t)=V(x,t+2\pi/\om)$ represents a classical
oscillatory field started at $t=0$. The latter is represented as a vector potential or, in the
length gauge, a dipole field, e.g. $V(x,t)=e\,E\cdot x\, \sin\om t$ \cite{e},\,\cite{o1}. One
then considers the time-dependent solution $\psi(x,t)$ of the Schr\"odinger
equation for $t>0$ starting with an initial state $\psi(x,0)=u_b(x)$, a bound state of $H_0$ with energy $-E_b$. 
$\psi(x,t)$ can be represented as a superposition of the initial bound state and of the generalized eigenstates of $H_0$ with momentum $\hbar k$, $u(k,x)$ i.e. asymptotically free scattering states: 
\begin{equation}\label{psi}
\psi(x,t)= \theta(t)e^{iE_bt}u_b(x)+\int_{\RR^d} \Theta(k,t)u(k,x)e^{-i\hbar^2k^2/2m}\, dk
\end{equation}
where we have assumed that there is only one (relevant) bound state.

In \eqref{psi} $|\theta(t)|^2$ is the probability at time $t$ that the particle is still in its bound state and $|\Theta(k,t)|^2$ is the probability density of finding the ionized electron in the (quasi) free state with energy $\hbar^2k^2/2m$. The unitarity of the evolution then gives
 $|\theta(t)|^2+\int_{\RR^d} |\Theta(k,t)|^2 dk=1$.

When $\hbar\om>E_b$, first order perturbation theory in the strength of $V$ (used very judiciously) yields, for "long times", emission into states $u(k,x)$ with $\hbar^2k^2/2m+E_b=\hbar\om$. This is interpreted as representing the absorption of one photon even though it is known that perturbation theory is not valid for "very long times" \cite{e}, \cite{o1}; see also \cite{CS}, \cite{SW}.
The clever use of first order perturbation theory also yields Fermi's Golden Rule of exponential decay of $|\theta(t)|^2$ from the initial bound state. To deal with the case of transitions caused by "$n$ photons", which one observes as "peaks" in noisy emission data \cite{b1}, one needs in principle to go to $n$'th order perturbation theory. This is very complicated and unreliable, so it is almost never attempted in practice. Instead one uses the so called strong field approximation due to Keldysh and others \cite{f},\,\cite{delta2},\,\cite{Proto},\,\cite{b1}. These are basically uncontrolled approximations which however give qualitative good results.

{\bf Description of the model.} To gain a clearer  picture of how resonances from a time-periodic potential give rise to peaks in the emitted energy distributions which look similar to $n$ 
photon absorption,  it is desirable to obtain an exact solution of the Schr\"odinger equation for arbitrary $t,\, \omega$ and strength of $V$.  
 In this work we describe exact results for $\psi(x,t)$, $\theta(t)$ and $\Theta(k,t)$ for a very simple 1d model system with $
H_0=-\frac{\partial^2}{\partial x^2}-2\delta(x)$ and $V(x,t)=-2\alpha \sin(\omega t)\delta(x)$. (We are using units in which $\hbar=2m=1$.)

The Hamiltonian $H_0$ has a single bound state $
u_b(x)=e^{-|x|},\ \ x\in\RR$, 
with energy $-E_b=-1$, and continuum states
\begin{equation}\label{uk}
u(k,x)=\frac{1}{\sqrt{2\pi}} \left( e^{ikx}-\frac{ e^{i|kx|}}{1+ik}  \right),\ \ x,k\in\RR
\end{equation}

 We then look for solutions of the time dependent Schr\"odinger equation
 
\begin{equation}\label{eqpsi}
i\psi_t=-\psi_{xx}-2\left(1+\alpha \sin\omega t\right)\,\delta(x)\,\psi
\end{equation}
for all positive $\alpha,\,\omega$ and $t$. 

This model has been studied extensively before  \cite{g},\,\cite{m},\,\cite{delta1},\,\cite{o8}, see in 
particular \cite{h} and references there. There $\theta(t)$ was proven to go to zero as $t\to\infty$. Its form for small $\alpha$ was obtained by a combination of analytic results and numerics, and shown to have many features similar to those obtained experimentally for the ionization of hydrogen-like atoms in a microwave electric field. There was however no computation of $\Theta(k,t)$ and $\psi(x,t)$. In this paper we present the physical content of new results for these quantities. These show, for the first time we believe, resonances in the energy distribution $|\Theta(k,t)|^2$ which correspond to multiphoton absorption for very long times. These are based on exact expressions in the form of multi-instanton expansions \cite{i}. We also present a new computation of the space-time structure of the wave function. Finally, we describe new, efficient numerical methods which we believe will be useful for more realistic systems.
(For numerical studies see \cite{g}, \cite{o8} and \cite{o1}.)

{\bf Results for $\Theta(k,t),\,\theta(t)$.}
 It was shown in \cite{h} that

\begin{equation}
  \label{eq:theta1}
  \theta(t)=1+2i\int_0^t \phi(s)ds
\end{equation}
and 
\begin{equation}
  \label{eq:theta1T}
 \Theta(k,t)=\sqrt{\frac{2}{\pi}}\frac{|k|}{1-i|k|}\,\int_0^t\phi(s)\, \mathrm{e}^{i(1+k^2)s}\, ds
\end{equation}
where $\phi$ satisfies the integral equation
$$\phi(t)=\alpha \sin\omega t\left(1+\int_0^t \phi(s)\eta(t-s)ds\right)$$
with
$$\eta(s)=i \left(\text{erf}\left(\sqrt{-i s}\right)+\frac{e^{i s}}{\sqrt{\pi } \sqrt{-i
   s}}+1\right)$$
It can be checked, \cite{h},  that the Laplace transform of $\phi$
\begin{equation}\label{star}
F(p):=\mathcal{L}\phi(p)=\int_0^{\infty}\phi(s)e^{-ps}ds
\end{equation}
is analytic in the right half plane and satisfies the functional equation
\begin{equation}
  \label{eq7}
  F(p)=\frac{i\alpha  }2\frac{F(p-i\omega)}{i\sqrt{ip+\omega-1}+1} -\frac{i\alpha  }2\frac{F(p+i\omega)}{i\sqrt{ip-\omega-1}+1} +\frac{\alpha \omega}{\omega^2+p^2}
\end{equation}
(The square root is understood to be positive on $\RR^+$, and analytically continued on its Riemann surface.)

In order to carry out the analysis of eq. \eqref{eq7} we define the function $\Phi(q):=F(-iq)$. This is analytic in $q$ in the closed upper half plane except for branch points at $q=n\omega+1$, and in the lower half plane except for an $\omega$-spaced array of poles parallel to the real line--representing physically resonances. The two step recurrence satisfied by $\Phi(q_0+n\omega)$ can be solved efficiently by continued fractions, or even more rapidly by a doubling procedure described in the sequel.

One can rewrite $\theta(t)$ and $\Theta(k,t)$ as Fourier transforms of functions with the same analyticity properties as $\Phi$. It was shown in \cite{i} that the functions $\Theta(k,t)$ and $\theta(t)$ have Borel summable multi-instanton expansions in $1/t$ valid for all $t>0$, for all $\alpha,\omega$. These explicit formulas are essentatially obtained by pushing the Fourier contour in the lower half plane, collecting residued (resulting in small exponentials) and Hankel countours around branch cuts, which are in fact Borel sums of asymptotic series in powers of $1/t$. For  small $\alpha$, $\Theta(k,t)$ and $\theta(t)$  have exact expressions in $\alpha$ for all $t$, provided that certain sums in the exponentials are kept in the exponent. On the other hand, a pure power series expansion in $\alpha$, as used in classical perturbation theory,  converges only for $t$ up to order $\alpha^{-1}\log \alpha$.

The Fourier transform interpretation  also provides a good intuition into the time behavior of the model. For small $\alpha$, with $m$ the least integer s.t. $m\omega>1$ the poles  are at a distance $O(\alpha^{2m})$  from the real line.  The largest residue  is  $O(\alpha^{2m})$.  This gives  Lorenzian shaped peaks in the profile of $\Phi$, of amplitude  $O(\alpha^{-2m})$. The total area under the graph of $|\Phi|$ comes solely from the peaks, up to $O(\alpha^{2m})$  corrections.  When $\omega^{-1}$ is not too close to an integer, $\Phi$ is analytic in a neighborhood of these peaks.  Its Fourier transform  decays exponentially with exponential rate $O(\alpha^{2m})$  up to times of order $\alpha^{-2m}$  when the oscillation in the Fourier transform is fast enough to cancel the peaks. Up to that time,  $|\theta(t)|^2= e^{-2\alpha^2m  \gamma(\omega) t}(1+o(1))$ (where $\gamma$ is $\alpha$-independent)\footnote{This is the Fermi Golden Rule of exponential decay of the bound state valid for small amplitudes and moderately large times.}. From $t\sim \alpha^{-2m}$  on, the much slower decay associated to the branch cuts, $O(\alpha ^4 t^{-3})$ takes over.  (The exact expression in the form of a convergent multiinstanton expansion leads to the same conclusion.) When however   $\omega^{-1}$ is close to an integer, $\Phi$ has cusps at the peaks, and the decay is not exponential anymore, not even on shorter time scales; instead it is   $O(\alpha ^4 t^{-3})$ mixed with high amplitude oscillations, at all times. Ionization is then much slower, a form of stabilization in this and similar models. This qualitative reasoning works even when $\alpha$ is of order one, in which case, due to the Stark shift, some other, $\alpha$-dependent values of $\omega$ result in cusped peaks.

It was also shown in \cite{i} that  $\lim_{t\to\infty}\Theta(k,t):=   \Theta(k,\infty)$ exists. For small $\alpha$ and $\omega>1$ it has the Lorentzian  shape:
  \begin{equation}
   \label{eq:ThetaSmall a}
   \Theta(k,t)=\sqrt{\frac2{\pi}}\,\,\frac{\alpha \omega k}{1-ik}\,\,\frac{1-e^{-\frac{\alpha  ^2 t \sqrt{\omega -1}}{2 \omega }} e^{i t \left(\frac{\alpha  ^2
   \sqrt{\omega +1}}{2 \omega }+k^2-\omega +1\right)}}{\alpha  ^2 \left(\sqrt{\omega -1}-i
   \sqrt{\omega +1}\right)-2 i \omega  \left(k^2-\omega +1\right)}(1+o(\alpha;t))
\end{equation}

\begin{figure}
  \includegraphics[width=0.7
  \textwidth, angle = 0]{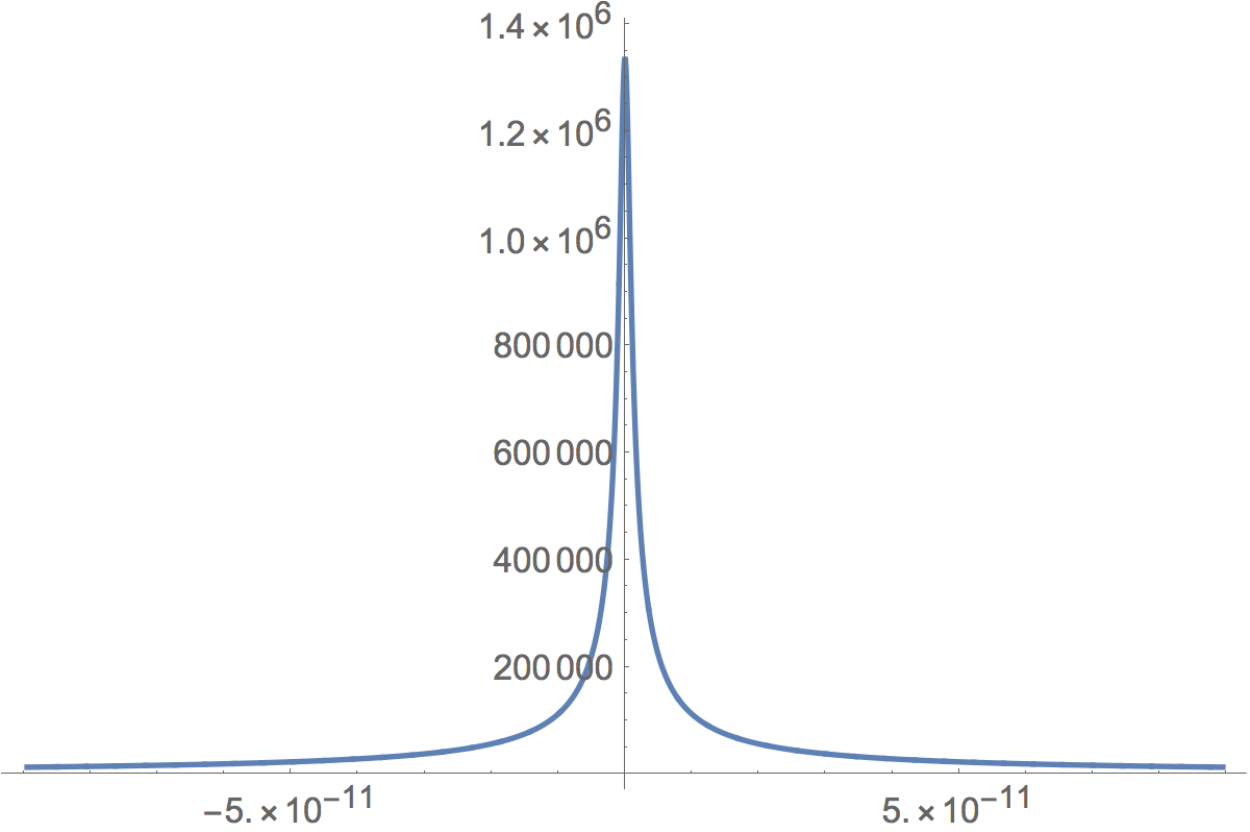} 
\caption{ $|\Theta(k_0+x,\infty)|$ for $k_0=0.1999489220447,\omega=0.4,\alpha=0.01$.}
\label{fig1AA}
 \end{figure}
 
 \begin{figure}
\includegraphics[width=0.7\textwidth, angle = 0]{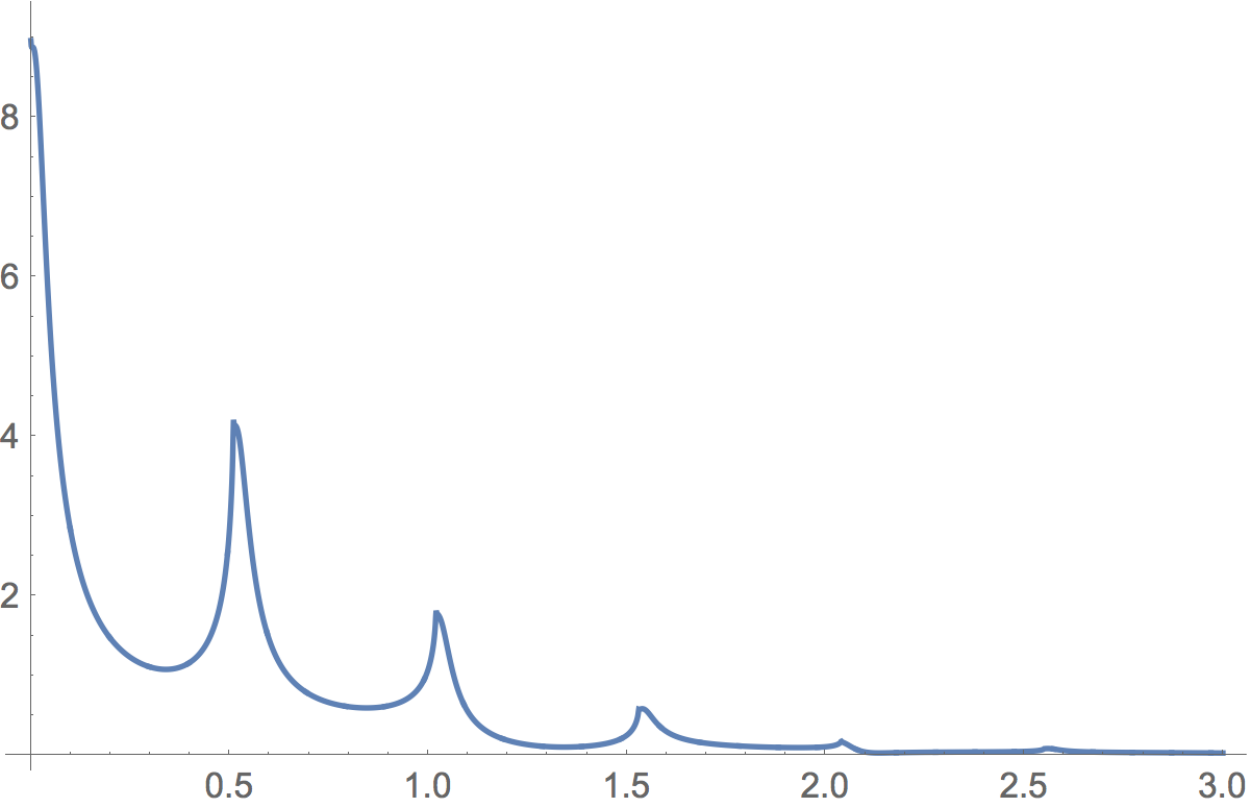}
\caption{\ \ \ \ $|\Theta(k,\infty)|$ for  $\alpha=1,\omega= 0.51$.}
\label{fig1B}
\end{figure}

 \begin{figure}
\includegraphics[width=0.7\textwidth, angle = 0]{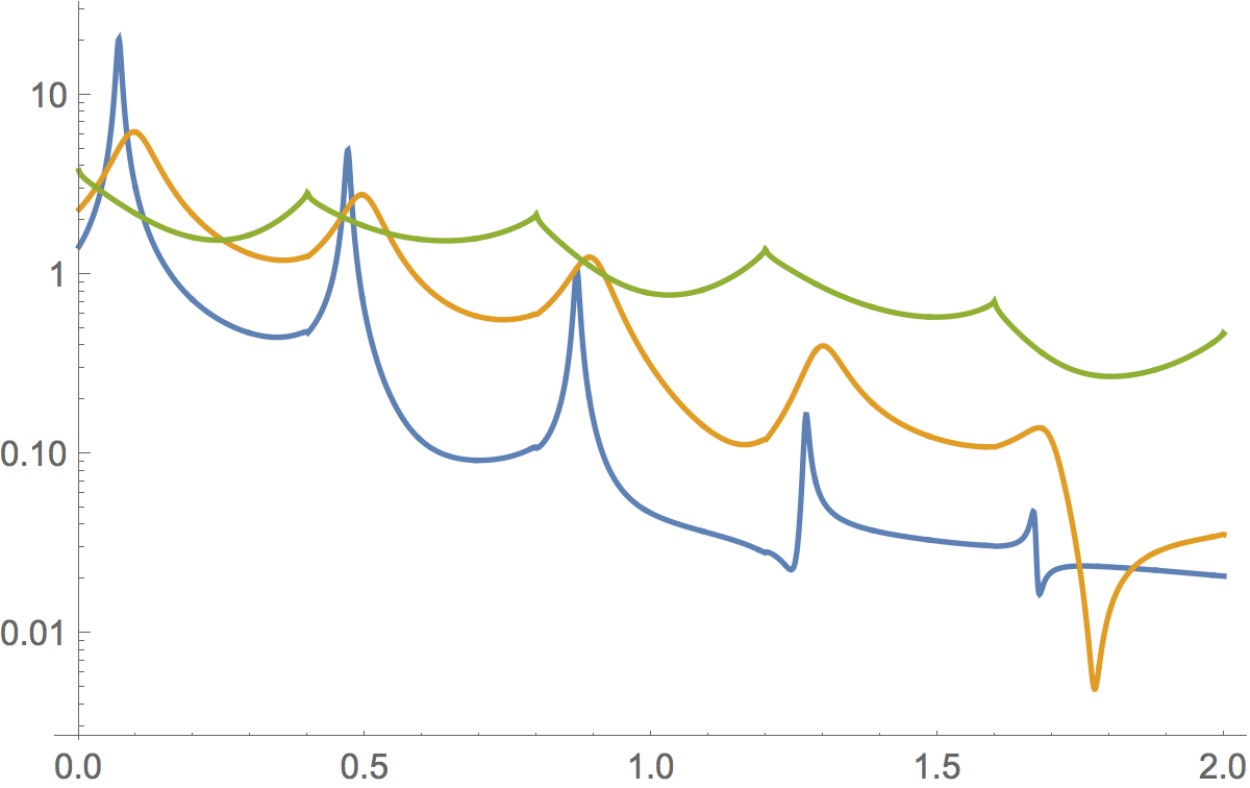}
\caption{Log plots of  $|\Theta(k,\infty)|$ for $\omega=0.4$ $\alpha=1/2,1, 2$. }
\label{fig1BB}
 \end{figure}

 \begin{figure}
  \includegraphics[angle=270,origin=c,width=0.5\textwidth, angle =90]{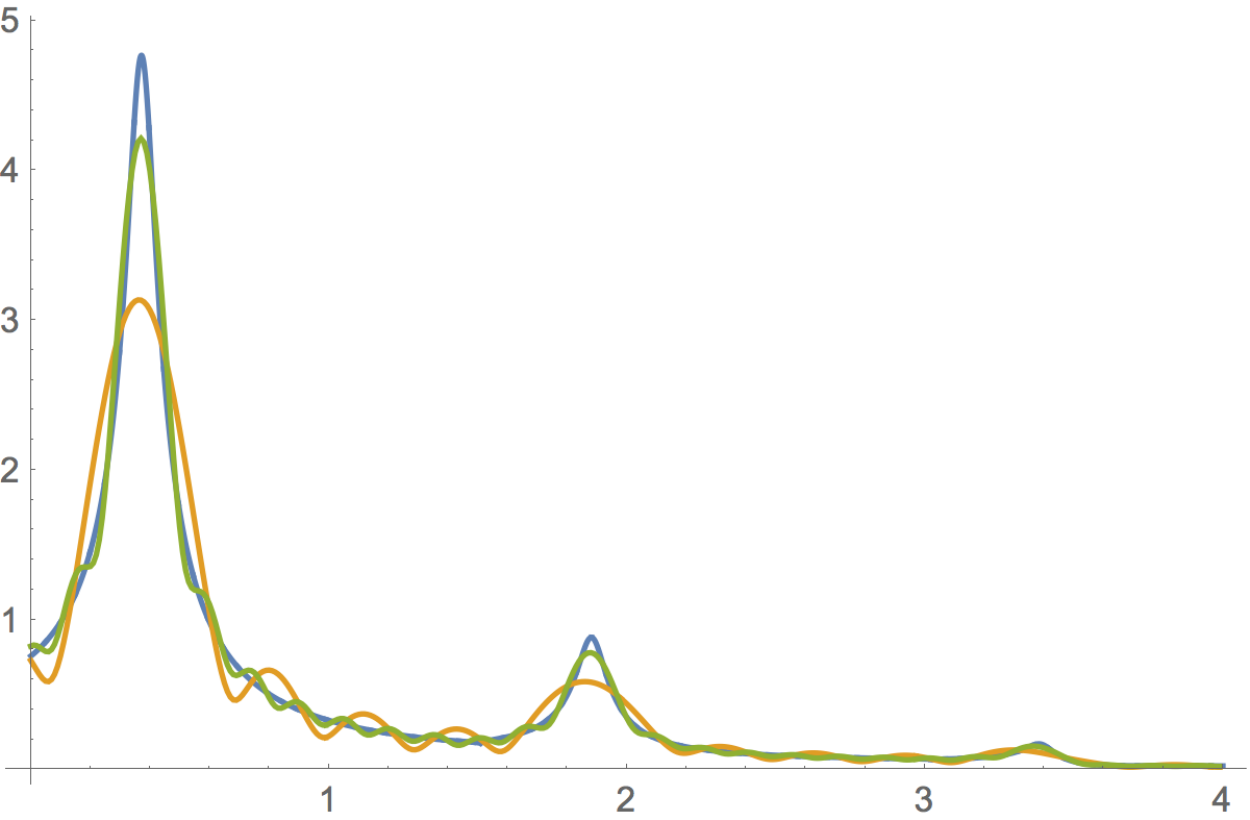}
\caption{Plot of $|\Theta(k,t)|^2$ at $\omega=1.51$, $\alpha=0.5$ and  $t=5T,10T,\infty$, $T:=\frac{2\pi}{\omega}$. The absolute maximum increases in this order.    }
\label{fig4BB}
\end{figure}


From \eqref{eq:ThetaSmall a} it follows that, for $\alpha\to 0$, after $t\to\infty$, $\Theta$ becomes a delta
function at $k^2= \omega -1$. There is a similar behavior for $\omega<1$, the delta function now occurring at $k^2=m\omega-1$, where $m$ is (as above) the smallest integer so that $m\omega>1$,
 see Fig. \ref{fig1AA} where we display the very sharp peak for $\alpha=.01,\omega=.4$ and $m=3$. There is a (Stark) shift of order $\alpha^2=10^{-4}$. The shape is close to a Lorentzian.

  The peaks in $|\Theta(k,t)|^2$ remain sharp for small $\alpha$
  and are located close to energies $k^2=n\omega -1,\ n\Ge m$. In Fig.\,\ref{fig1B} we show $|\Theta(k,t)|$ for $\omega=.51,\ \alpha=1$. The peak at zero corresponds to 3-''photon'' ionization (because of the Stark shift; 2 would be needed at small $\alpha$). One can distinguish peaks corresponding to up  8-photon ionization. The fact that $\alpha$ is large permits us to see that many peaks. For small values of $\alpha$ the peaks for $(n+1)\omega$ are smaller than those for $n\om$ by a factor of $\alpha^2$, see Fig. \ref{fig1BB} where one can also see how the peaks broaden when $\alpha$ gets large: they essentially disappear for $\alpha \gtrsim 3$.
  For finite times the peaks broaden and get smaller. They are however still
  visible when $t$ is of the order of a few periods, see Fig.\,\ref{fig4BB}.  The reason for this is that $\Theta(k,t)-\Theta(k,\infty)$ decays in a  manner similar to that of $\theta$. This explains why the limiting profile is visible after only a few oscillations if $\alpha$ is not extremely small. 
  
   \begin{figure}
 \includegraphics[width=0.7\textwidth, angle = 0]{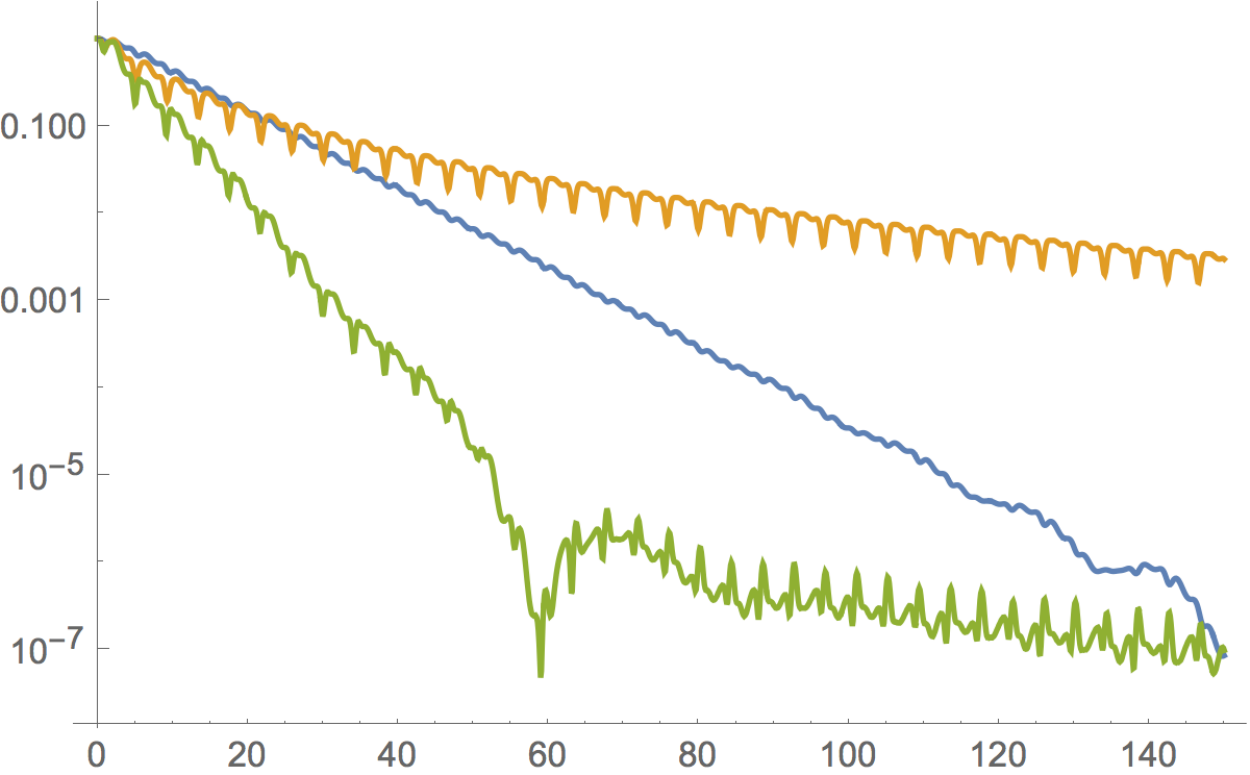}
 \caption{Log plot of $|\theta(t)|^2$ for $\omega=1.51$ at $\alpha=0.5$ (almost straight line);  at $\alpha=0.98$, yellow.; $\alpha=1.3$ (lowest curve).  }
\label{fig4AA}
\end{figure}

    In Fig.\,\ref{fig4AA} we can see several graphs of $|\theta(t)|^2$, the survival probability versus $t$. For $\alpha=1/2$ the Fermi Golden rule is clearly visible, for all relevant times. At larger times, the $t^{-3}$ behavior kicks in, again mixed with oscillations, as predicted by the multi-instanton expansions: at $\alpha=.98$ we observe stabilization: the decay is always power-like and slower, except at very short times. This is consistent with the stabilization process that we mentioned. At $\alpha=1.3$ the log-plot shows an initial exponential behavior, followed by rough dip when the exponential and polynomial parts become comparable, resulting in some cancellations, after which the behavior becomes polynomial, with oscillations.

All the above results do not use perturbation theory but agree with it when the latter is applicable. We obtained the plots of $\Theta$ and $\psi$  by numerically taking the inverse time Laplace transform of $\psi(x,t)$ for moderate time, and then by stationary phase calculation of the inverse Laplace transforms. Various features of $|\Theta(k,t)|^2$ are similar to those
  observed in experiments \cite{b1}.

A more direct connection between this semi-classical description and the "photon" picture can be made via Floquet theory \cite{g}. Using a suitable representation for the laser field in  a cavity one can describe the absorption of $n$ (non-localized) photons by an atom in terms of the solution of a Schr\"odinger equation. We shall consider the connection between our results and this formalism in a future work \cite{k}.


{\bf Calculation of $\psi(x,t)$.}
 The full behavior of $\psi(x,t)$ is very complicated despite the simplicity of the model.  Here we present the main results of the new calculations, leaving the details for another paper \cite{l}. We expect the main feature of the evolution of $\psi(x,t)$ to be universal for ionization by an oscillatory field \cite{m}. 
 
\begin{figure}
  \includegraphics[width=0.7\textwidth, angle = 0]{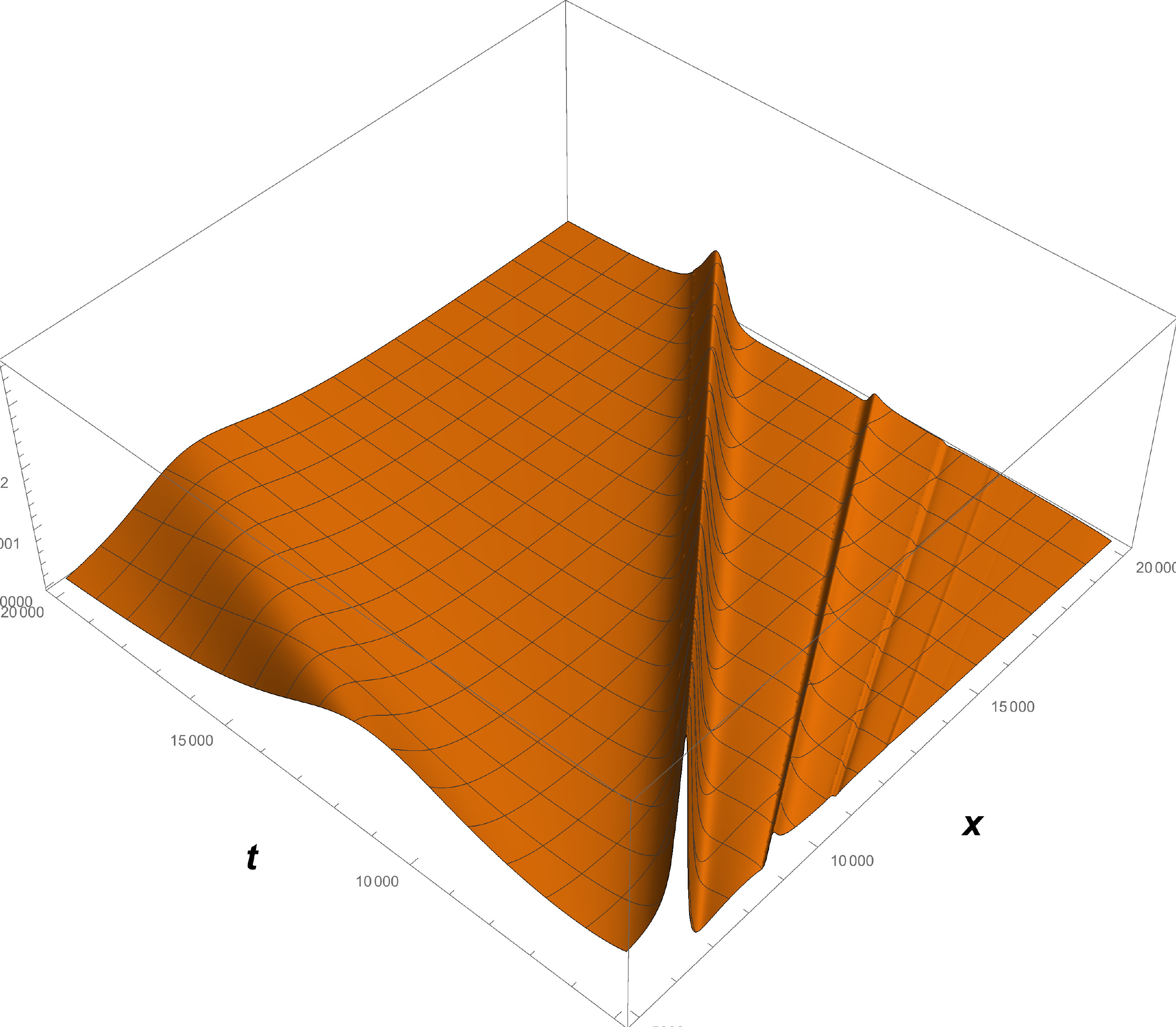}
\caption{ $|\psi(x,t)|^2$ for $\omega= 1.52,\alpha=1.5$ }
\label{fig6}
\end{figure}

    For
small values of $x$ and $t$, $|\psi(x,t)|$ is highly oscillatory, indicating the formation of
wave packets. For large $x$ and $t$  of comparable order of magnitude, the behavior
suggests trajectories of free classical particles, as expected from the scaling
$t=N\tau$, $x=N\xi$, $N\gg 1$  which is roughly equivalent to taking
$\hbar\to\hbar/N$ and correspondingly suppressing the delta function. The
asymptotic behavior of $\psi$ in this regime is particularly simple, see
Fig. \ref{fig6}.

To obtain these results we use the Laplace transform $\hat{\psi}(x,p)$ defined in \cite{i}. Letting $\psi(x,0)=u_b(x)$,  equation \eqref{eqpsi} becomes
\begin{equation}\label{eqhpsi}
-i u_b+ip\hat{\psi}=-\hat{\psi}_{xx}+ R(p)\delta(x)
\end{equation}
where 
\begin{equation}\label{formR}
R(p)=-2\hat{\psi}(0,p)+i\alpha\left[\hat{\psi}(0,p-i\omega)-\hat{\psi}(0,p+i\omega)\right]
\end{equation}

For $p$ in the right-half plane any square-integrable solution of \eqref{eqhpsi} has the form
\begin{equation}\label{solhpsi}
\hat{\psi}(x,p)=-\frac{R(p)}{2\sqrt{-ip}} \,e^{-|x|\sqrt{-ip}\, }+\phi(x,p)
\end{equation}
where $\Re \sqrt{-ip}>0$ and

\begin{equation}\label{phi}
\phi(x,p)=  \frac i{\sqrt{-ip}(-ip-1)}\,\left(e^{-|x|\sqrt{-ip}}-\sqrt{-ip}\,  e^{-|x|}\right)
\end{equation}

Combining \eqref{formR}-\eqref{phi} we obtain

\begin{equation}\label{recpsi}
\hat{\psi}(0,p)=\, -\frac{i\alpha}{2}\frac{1}{\sqrt{-ip}-1}\,\left[\hat{\psi}(0,p-i\omega)- \hat{\psi}(0,p+i\omega)\right] +\frac{i}{ip+1}
\end{equation}


To calculate $\hat{\psi}(0,p)$, denote
$$p=-iq,\ \ \ q=\sigma+n\omega\ \text{ where }\Re\sigma\in[0,\omega),\ \ \hat{\psi}(0,-i(\sigma+n\omega))=g_n$$
Equation \eqref{recpsi} now becomes
\begin{equation}\label{eqgn}
g_n=\, \frac{i\alpha}{2}\frac{1}{\sqrt{-\sigma-n\omega}-1}\,\left( g_{n-1}- g_{n+1}\right) +\frac{i}{\sigma+n\omega+1} 
\equiv a_n^{[0]} g_{n-1}+b_n^{[0]}g_{n+1} +f_n^{[0]}
\end{equation}
(where the square root is chosen so that $\sqrt{u}$ equals $-i\sqrt{|u|}$ if $u<0$). 

Methods similar to those of \cite{i} show that equation \eqref{eqgn} has a unique square-summable solution $g_n=g_n(\alpha,\sigma)$, analytic in $\alpha$ for small $\alpha$ and real analytic for all $\alpha$. It is also analytic in $\sigma$, except for a square root branch points at $0$ and for a pole of order one in the lower half plane; its residue can be calculated using a convergent continued fraction.

This solution of  \eqref{eqgn} can be calculated numerically, rapidly, with arbitrary precision, by iterating \eqref{eqgn} $N$ times, each time doubling the recurrence step, leading to
$g_n= a_n^{[N]} g_{n-2^N}+b_n^{[N]}g_{n+2^N} +f_n^{[N]}$
where now $a_n^{[N]},\,b_n^{[N]}=O(n^{-2^{N-1}})$. Using an $N$ large enough for the desired accuracy, one can then approximate  $g_n\approx f_n^{[N]}$ for all $n$ with $|n|<2^N$. For $|\alpha|<2$, $N= 6$ gives sufficient accuracy.

The Laplace transform of the wave function, \eqref{solhpsi}, satisfies

\begin{equation}\label{dechp}
\hat{\psi}(x,p)=e^{-|x|\sqrt{-ip}}\hat{\psi}(0,p)+ \frac{ie^{-|x|\sqrt{-ip}}}{\sqrt{-ip}(1+\sqrt{-ip})} + \frac {i\left(e^{-|x|\sqrt{-ip}}-\sqrt{-ip}\,  e^{-|x|}\right)}{\sqrt{-ip}(-ip-1)}
\end{equation}

This yields for large $x\sim t$ the simple formula

\begin{equation}\label{star}
\psi(x,t)\sim e^{i\frac{x^2}{4t}}\,  \frac 1{2\sqrt{i\pi}}\,   \frac{|v|}{\sqrt{t}}\, \left[\hat{\psi}\left(0,-iv^2/4\right)  -\frac{i}{1+v^2/4} 
\right],\ \ \ \text{where }v=\frac xt=O(1)
\end{equation}
where $\hat{\psi}(0,p)$ is given explicitly as a convergent continuous fraction. In this semiclassical limit we have $E=\frac{mv^2}{2}$, $p=mv$, $E=k^2=n\omega -1$; the "trajectories" in Fig. \ref{fig6}  reflect this correspondence.

We also obtained formulas for $\psi(x,t)$ for moderate $x$ and $t$.

\section{Acknowledgments}
OC  was partially supported by the NSF-DMS grant 1515755 and JLL by the AFOSR grant FA9550-16-1-0037. We thank H. Jauslin, H. Spohn and particularly L. diMauro, C. Blaga and David Huse for very useful discussions. JLL thanks the Systems Biology division of the Institute for Advanced Study for hospitality during part of this work.


\begin{thebibliography}{99}

\bibitem{e} D Bauer, 2006 {\em Theory of intense laser-matter interaction}, Lecture notes, Univ. of Heidelberg

\bibitem{delta1} W Becker, S Long and JK McIver  1994, {\em Modeling harmonic generation by a zero-range potential}, Phys. Rev. A 50 1540



\bibitem{b2} M F Ciappina et. al, 2017 {\em Attosecond physics at the nanoscale}, Rep. Prog. Phys. 80 054401

\bibitem{c} Cohen-Tannoudji C., Dupont-Roc J., Grynberg G., Cohen-Tannoudji C., Dupont-Roc J., Grynberg G., {\em Photons and Atoms. Introduction to quantum electrodynamics.}, 1997 {\em Photons and Atoms. Introduction to quantum electrodynamics.}, John Wiley and Sons Ltd

\bibitem{o1} Cohen-Tannoudji C, Duport-Roc J and Arynberg G 1992 {\em Atom-Photon Interactions}, (NewYork:Wiley) Chin S L and Lambropoulus P (ed) 1984 Multiphoton Ionization of Atoms (New York: Academic)



\bibitem{i} O Costin, R D Costin, J L Lebowitz, {\em Nonperturbative time dependent solution of a simple ionization model.} arXiv:1706.07129 

\bibitem{k} O Costin, R Costin, H Joselin, I Joselin, JL Lebowitz, {\em in preparation}

\bibitem{l} O Costin, R Costin, JL Lebowitz, {\em in preparation}

\bibitem{h} O Costin, J L  Lebowitz, A Rokhlenko, 2000 {\em Exact Results for the Ionization of a Model Quantum System} J. Phys. A: Math. Gen.  33 pp. 1--9 



\bibitem{CS} O Costin, A. Soffer 2001 {\em Resonance Theory for Schr\"odinger Operators}, Commun. Math. Phys. 224 


\bibitem{a} N B Delone and V P Krainov, {\em Multiphoton Processes in Atoms}, Springer, Berlin Heidelberg, New York, 1994 



\bibitem{g} Demkov Yu N and Ostrovskii V N 1988 {\em Zero Range Potentials and Their Application in Atomic Physics} (New
York: Plenum)

\bibitem{m}  W. Elberfeld and M. Kleber, {\em Tunneling from an ultrathin quantum well
in a strong electrostatic field: A comparison of different methods.} Z. Phys.B-- Condensed Matter 73, 23--32 (1988). 


\bibitem{delta2} FHM Faisal 1973 {\em Collision of electrons with laser photons in a background potential}, J. Phys. B: At. Mol. Phys. 6 No.11, L312


\bibitem{j} Gu\'erin S., Monti F., Dupont J-M., Jauslin H. R., {\em On the relation between cavity-dressed states, Floquet states, RWA and semiclassical models}, J. Phys. A, 30 (1997)


\bibitem{f} S V Popruzhenko 2014 {\em Keldysh theory of strong field ionization: history, applications, difficulties and
perspectives}, J. Phys. B: Atomic, Molecular and Optical Physics 47 (20) 

\bibitem{Proto} M Protopapas et al 1997 {\em Atomic physics with super-high intensity lasers}, Rep. Prog. Phys. 60 389

\bibitem{delta3} Reiss H R 1980 {\em Effect of an intense electromagnetic field on a weakly bound system}, Phys. Rev. A 22 1786

\bibitem{b1} Schenk M, Kr\"uger M and Hommelhoff P 2010 {\em Strong-field above-threshold photoemission from sharp metal tips}, Phys. Rev. Lett. 105 257601


\bibitem{SW} A. Soffer, M. Weinstein 1998 {\em Time Dependent resonance theory}, Geometric and Functional
Analysis, (GAFA) 8, 1086-1128.



\bibitem{d} H Spohn 1991 {\em Dynamics of Charged Particles and their Radiation Field}, Cambridge University Press, 2004



\bibitem{o8} SM Susskind, SC Cowley and EJ Valeo 1994 {\em  Multiphoton ionization in a short range potential: A nonperturbative approach}, Phys.Rev. A 42 3090 








 \end{thebibliography}
\end{document}